\documentclass{ws-procs10x7}

\begin{document}

\title{Track Reconstruction and Alignment with the CMS Silicon Tracker}

\author{F.-P. SCHILLING (CMS Collaboration)}

\address{
Physics Department, CERN, CH-1211 Geneva 23, Switzerland\\
E-mail: frank-peter.schilling@cern.ch
}

\twocolumn[\maketitle\abstract{
This report presents recent results on track reconstruction
and alignment with the silicon tracker of the CMS experiment at the
LHC, obtained with a full detector simulation. After an overview of
the layout of the tracker and its material budget, the baseline
algorithm for track reconstruction is discussed.  The performance of
the track reconstruction and its dependence on misalignment is shown.
The concept for alignment of the CMS tracker, using a laser alignment
system as well as three different track-based alignment algorithms, is
presented.
}]


\section{INTRODUCTION}

This report\footnote{Poster presented at ICHEP 2006, Moscow}
presents recent results~\cite{tdr} on track reconstruction
and alignment with the silicon tracker of the CMS experiment at the
LHC, obtained with a full detector simulation. After an overview of
the layout of the tracker and its material budget, the baseline
algorithm for track reconstruction is discussed.  The performance of
the track reconstruction and its dependence on misalignment is shown.
The concept for alignment of the CMS tracker, using a laser alignment
system as well as three different track-based alignment algorithms, is
presented.

\section{THE CMS SILICON TRACKER}

\begin{figure}[t]
\centering
\includegraphics[width=0.99\linewidth]{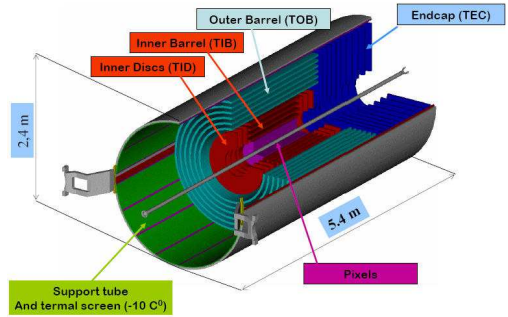}
\caption{Illustration of the CMS tracker. The various
components such as barrel and endcap strip and pixel detectors,
are housed in a support tube $2.4 \rm\ m$ in diameter and $5.4 \rm\ m$
in length.}
\label{fig:tracker}
\end{figure}

The CMS Silicon Tracker (Figure~\ref{fig:tracker}) is one of the main
components of the CMS experiment at the LHC. It consists of
$\sim15000$ silicon strip and pixel sensors covering an active area of
$\sim200 \rm\ m^2$ within the tracker volume of $24.4 \rm\ m^3$.  The
full tracker has a radius of $\sim110\rm\ cm$ and covers
pseudorapidity values up to $\eta=2.4$.

\begin{figure}[t]
\centering
\includegraphics[width=0.99\linewidth]{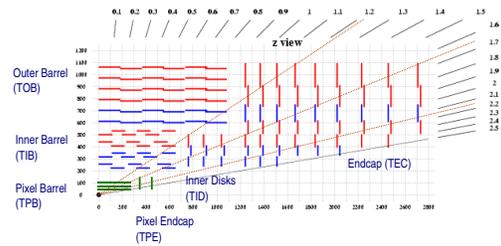}

\caption{Illustration of the CMS tracker layers, showing 
one quarter of the full tracker in $rz$ view.}
\label{fig:trackerlayers}
\end{figure}

The Barrel strip detector consists of 4 inner (TIB) and 6 outer (TOB)
layers (Figure~\ref{fig:trackerlayers}). The first two layers in TIB
and TOB use double-sided sensors. The Endcap strip detector is made of
3 inner (TID) and 9 outer (TEC) disks (rings 1,2 and 5 are double
sided). The Pixel detector consists of 3 barrel layers at $r=4.4, 7.3$
and $10.2 \rm\ cm$, and of two endcap disks.

\begin{figure}
\centering
\includegraphics[width=0.31\linewidth]{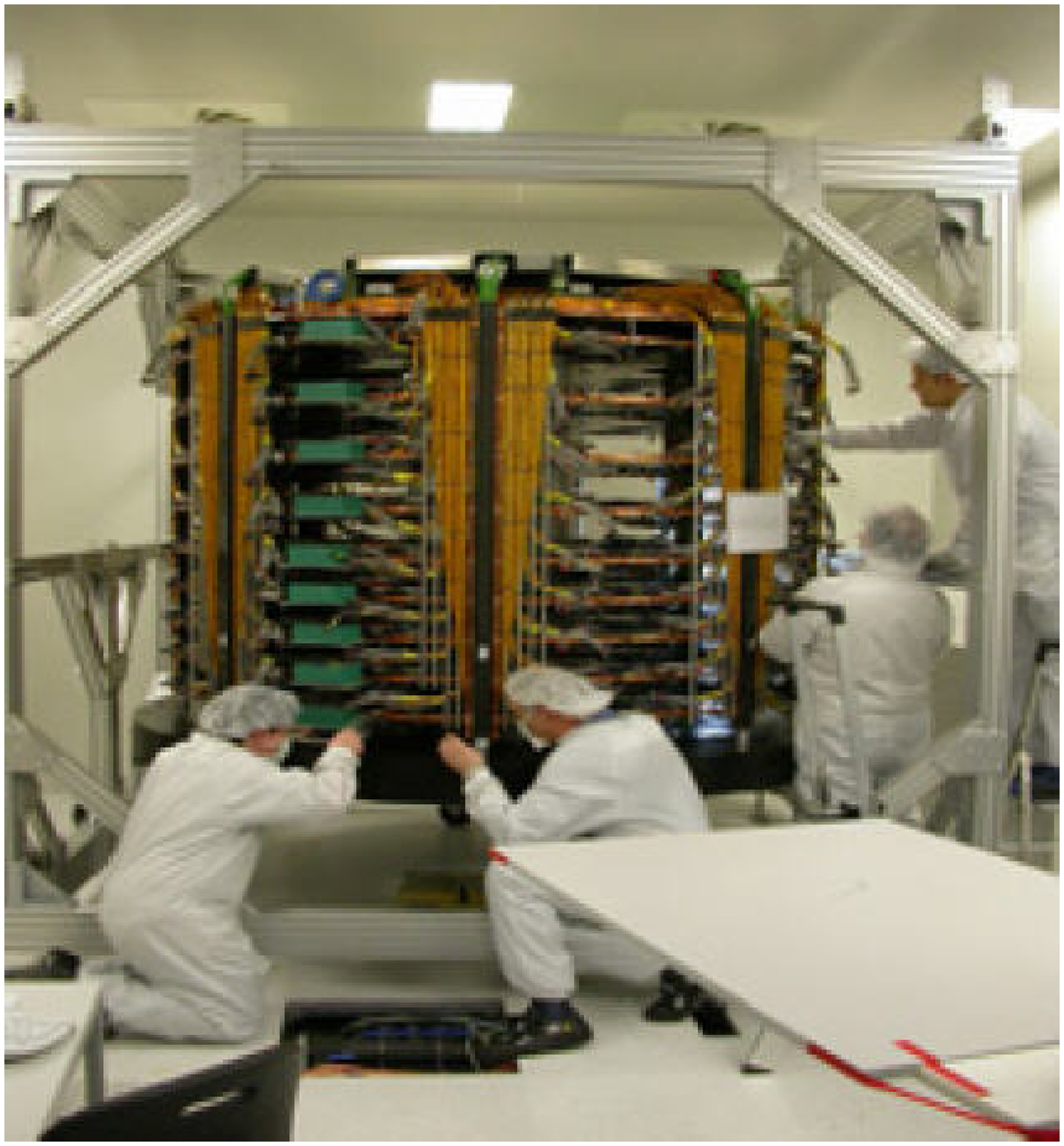}
\hspace{0.5cm}
\includegraphics[width=0.55\linewidth]{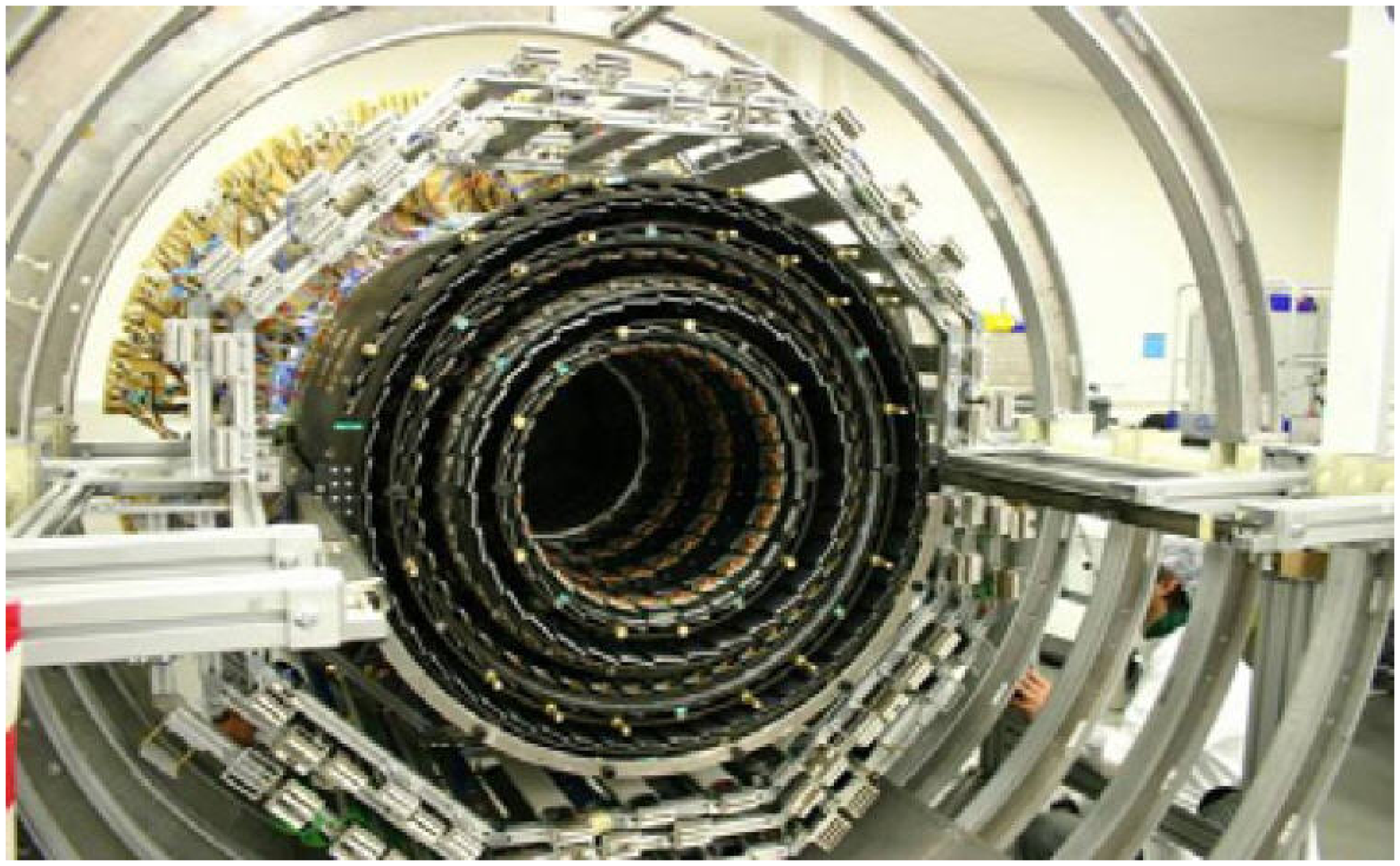}
\caption{Photographs from the integration of tracker components. Left: TEC Integration; Right: One half of TIB/TID completed.}
\label{fig:tidtec}
\end{figure}

The Strip Sensors consist of 512 or 768 strips with a pitch of
$80\ldots 200 \rm\ \mu m$ Their resolution in the precise coordinate
is in the range $20\ldots50 \rm\ \mu m$.  The Pixel sensors are made
of pixels of size $100 (r\phi) \ {\rm x} \ 150 (z) \rm\ \mu m^2$ with a
resolution of $10\ldots15 \rm\ \mu m$. The modules are mounted on
carbon-fiber structures and housed inside a temperature controlled
outer support tube. The operating temperature will be around $-20^{\rm
o} \rm C$. Figure~\ref{fig:tidtec} shows two recent photographs from
the integration of tracker components.

\section{TRACKER MATERIAL BUDGET}

The CMS tracker includes both sensitive volumes and non-sensitive
ones.  Since the tracker requires a large amount of low-voltage power,
a large amount of heat needs to be dissipated.  Therefore, a large
fraction of the tracker material consists of electrical cables and
cooling services. Other non-sensitive parts include support
structures, electronics, the beam-pipe, and the thermal screen outside
the tracker.

\begin{figure*}
\centering
\includegraphics[width=0.9\linewidth]{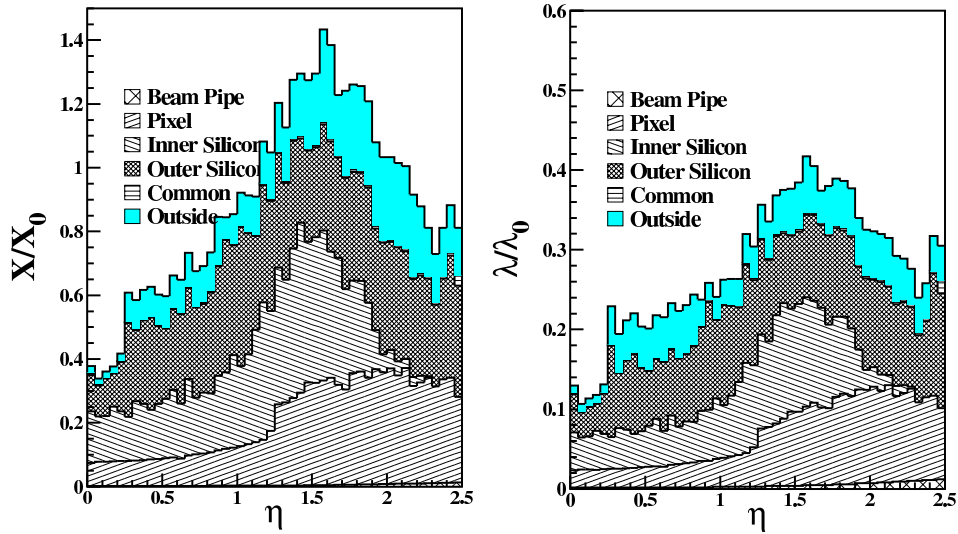}
\caption{Tracker material budget in units of radiation length (left) and interaction length (right) as a function of $\eta$ for the different subunits.}
\label{fig:material}
\end{figure*}

As a result, the tracker material budget can exceed the equivalent of
one radiation length for certain regions of $\eta$, which affects
hadron and electron reconstruction. The decomposition of the tracker
material in terms of radiation lengths and interaction lengths versus
$\eta$ for the different subdetectors is shown in
Figure~\ref{fig:material}.

\section{TRACK RECONSTRUCTION}

Track reconstruction in a dense environment needs an efficient search
for hits during the pattern recognition stage and a fast propagation
of trajectory candidates. In the CMS tracker, these tasks benefit from
the arrangement of the sensitive modules in practically hermetic
layers as well as from the almost constant four Tesla magnetic field
provided by the CMS solenoid magnet. Since the typical step length for
the propagation of track parameters is of the order of the distance
between two layers, a helical track model is adequate.

For reconstruction purposes the detailed distribution of passive
material as used in the simulation is replaced by an attribution of
material to layers. This model simplifies the estimation of energy
loss and multiple scattering, which can be done at the position of the
sensitive elements without additional propagation steps.

The baseline algorithm for track reconstruction~\cite{trackingnote} in
CMS is the Combinatorial Kalman Filter. After the tracker hits have
been reconstructed (clustering and position estimation), track
reconstruction proceeds through the following four stages:

\begin{itemize}
\item Trajectory Seeding
\item Pattern Recognition
\item Trajectory Cleaning
\item Track fitting and smoothing
\end{itemize}

In the following subsections, these steps are explained in 
more detail.

\subsection{Trajectory Seeding}

Seed generation provides initial trajectory candidates for the full
track reconstruction. A seed must define initial trajectory parameters
and errors. Hence, five parameters are needed to start trajectory
building. Therefore, the standard trajectory seeds in the CMS tracker
are constructed from pairs of hits in the pixel detector and a vertex
constraint. The pixel detector is well suited for seeding due to its
low occupancy, its proximity to the beam spot and due to the 2D
measurement capability in both $r\phi$ and $rz$. The seed finding
efficiency is $>99\%$.

Alternatively to the baseline seeding, a seeding using the innermost
layers of the strip tracker has also been implemented, to be used for
example at the start-up when the pixel detector will not yet be
installed. In addition, external seeds provided by the calorimeter or
the muon detector can be used.

\subsection{Pattern Recognition}

Trajectory building is based on a combinatorial Kalman filter
method. The filter proceeds iteratively from the seed layer, starting
from a coarse estimate of the track parameters provided by the seed,
and including the information of the successive detection layers one
by one. With each included layer, the track parameters are better
constrained.  In the extrapolation of the trajectory from layer to
layer, the effects of energy loss and multiple scattering are accounted
for.

Trajectory candidates are added for each compatible hit (including an
additional trajectory without a measured hit in order to account for
inefficiencies), and the trajectory parameters are updated according
to the Kalman filter formalism.  The best trajectory candidates are
grown in parallel up to the outermost layers.

\subsection{Trajectory Cleaning}

Ambiguities in track finding arise because a given track may be
reconstructed starting from different seeds, or because a given seed
may result in more than one trajectory candidate.  These ambiguities
must be resolved in order to avoid double counting of tracks.

The ambiguity resolution is based on the fraction of hits that are
shared between two trajectories. It is applied twice: the first time
on all trajectories resulting from a single seed, and the second time
on the complete set of track candidates from all seeds.

\subsection{Track fitting and smoothing}

For each trajectory, the building stage results in a collection of hits
and an estimate of the track parameters. However, the full information
is only available at the last hit of the trajectory, and the estimate
may be biased by constraints applied during the seeding
stage. Therefore the trajectory is refitted using a least squares
approach, implemented as a combination of a standard Kalman filter and
smoother.  While the filter runs inside-out, in the smoothing step a
second filter is run outside-in. In both cases, the initial covariance
matrix of the track parameters is scaled by a large factor to avoid
possible biases.  At each hit the updated parameters of the smoothing
filter are combined with the predicted parameters of the first
filter. The combination yields optimal estimates of the track
parameters at the surface of each hit.


\section{TRACKING PERFORMANCE}

\subsection{Track finding efficiency}

\begin{figure*}
\centering
\includegraphics[width=0.44\linewidth]{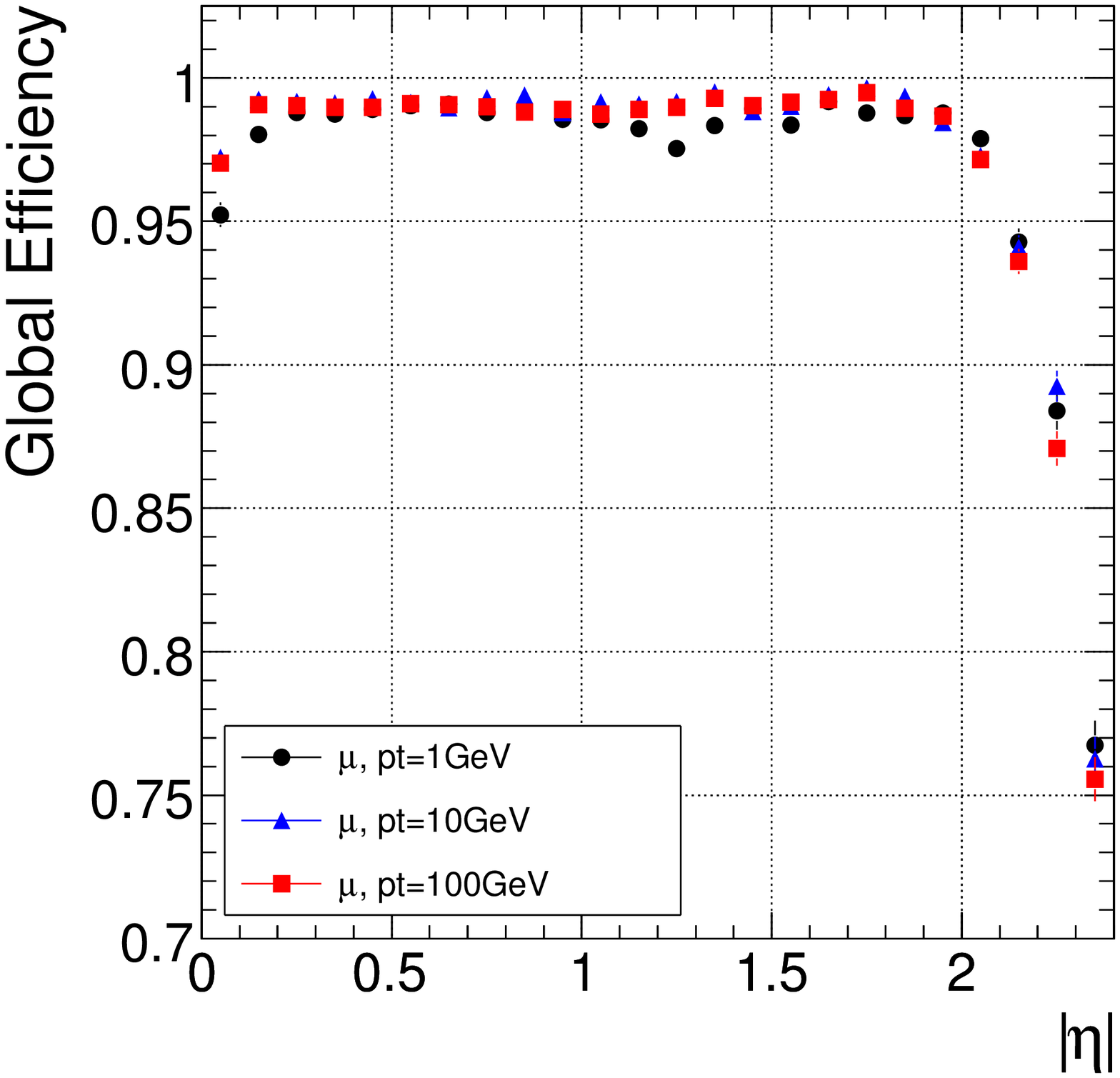}
\includegraphics[width=0.44\linewidth]{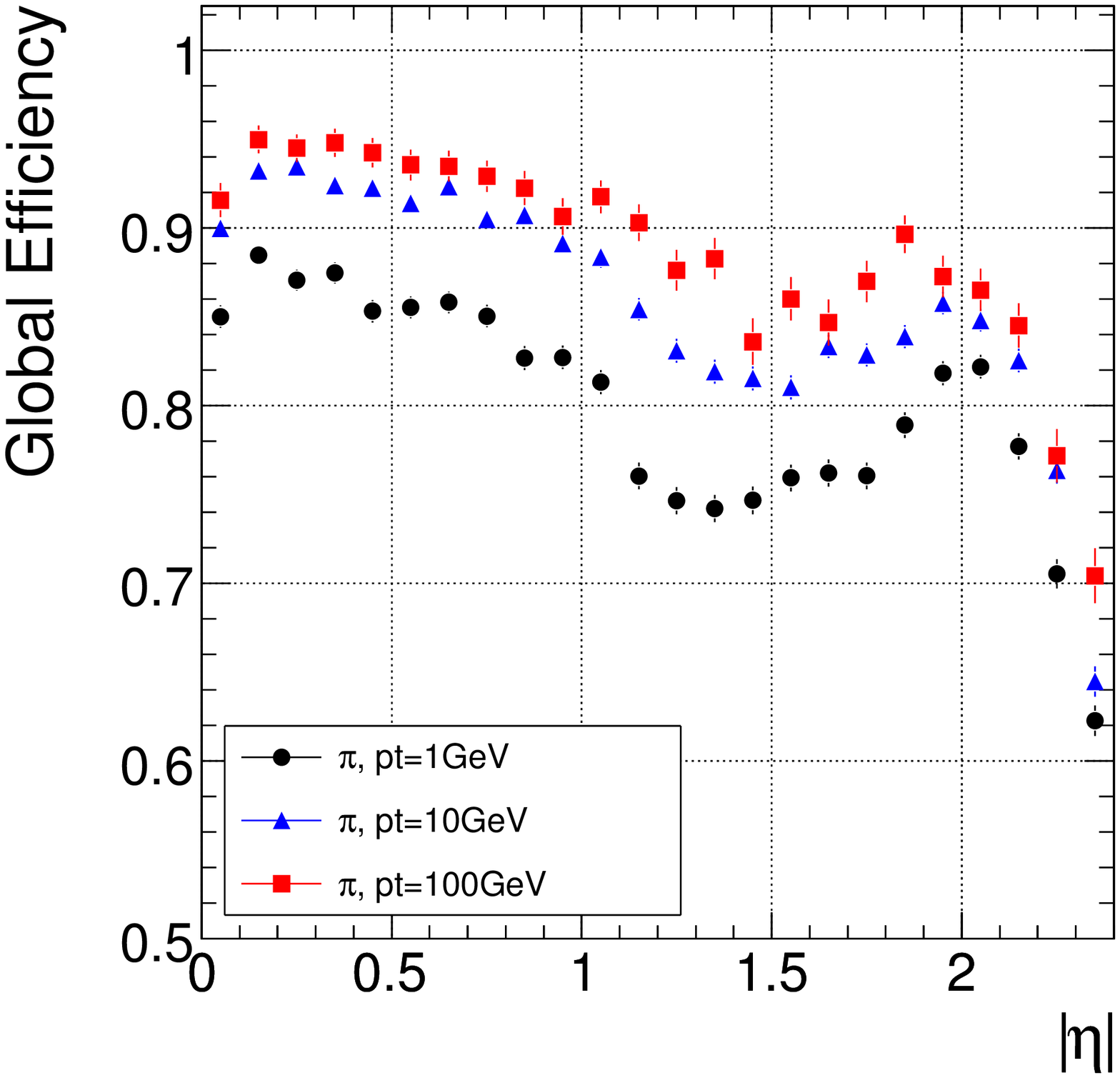}
\caption{Track finding efficiency for muons (left) and 
pions (right) with  $p_T=1,10$ and $100 \rm\ GeV$ as a function of  $\eta$.}
\label{fig:efficiency}
\end{figure*}

The efficiency for reconstructing single tracks with the combinatorial
Kalman filter has been estimated using samples of muons and pions with
transverse momenta of $1,10$ and $100 \rm\ GeV$. The results are shown
in Figure~\ref{fig:efficiency}. Here, reconstructed tracks are
required to have at least 8 hits and a minimum $p_T$ of $0.8 \rm\
GeV$. A track is deemed to be successfully reconstructed if it shares
more than $50\%$ of the hits with a simulated track.

The global track finding efficiency for muons is excellent, exceeding
98\% over most of the tracker acceptance. The drop of efficiency in
the region $|\eta|<0.1$ is due to the gaps between the sensors in the
ladders of the pixel detector at $z=0$. At high $\eta$, the drop in
efficiency is mainly due to the lack of coverage by the two pairs of
pixel endcap disks.

For hadrons, the efficiency is between $75$ and $95\%$, depending on
momentum and $\eta$. It is lower compared with the efficiency for muons
because the hadrons interact with the tracker material.

\subsection{Resolution}

\begin{figure*}
\centering
\includegraphics[width=0.44\linewidth]{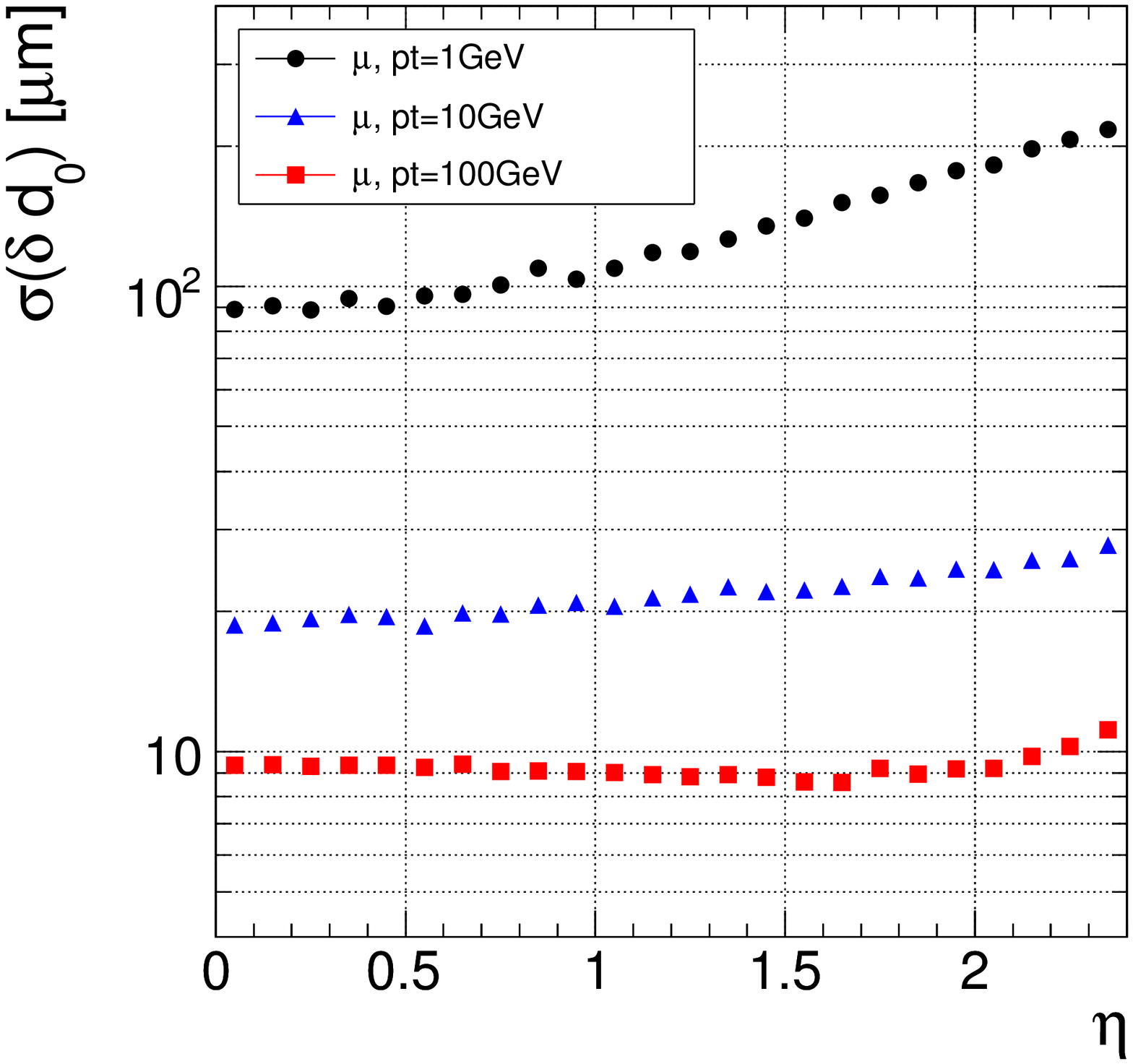}
\includegraphics[width=0.44\linewidth]{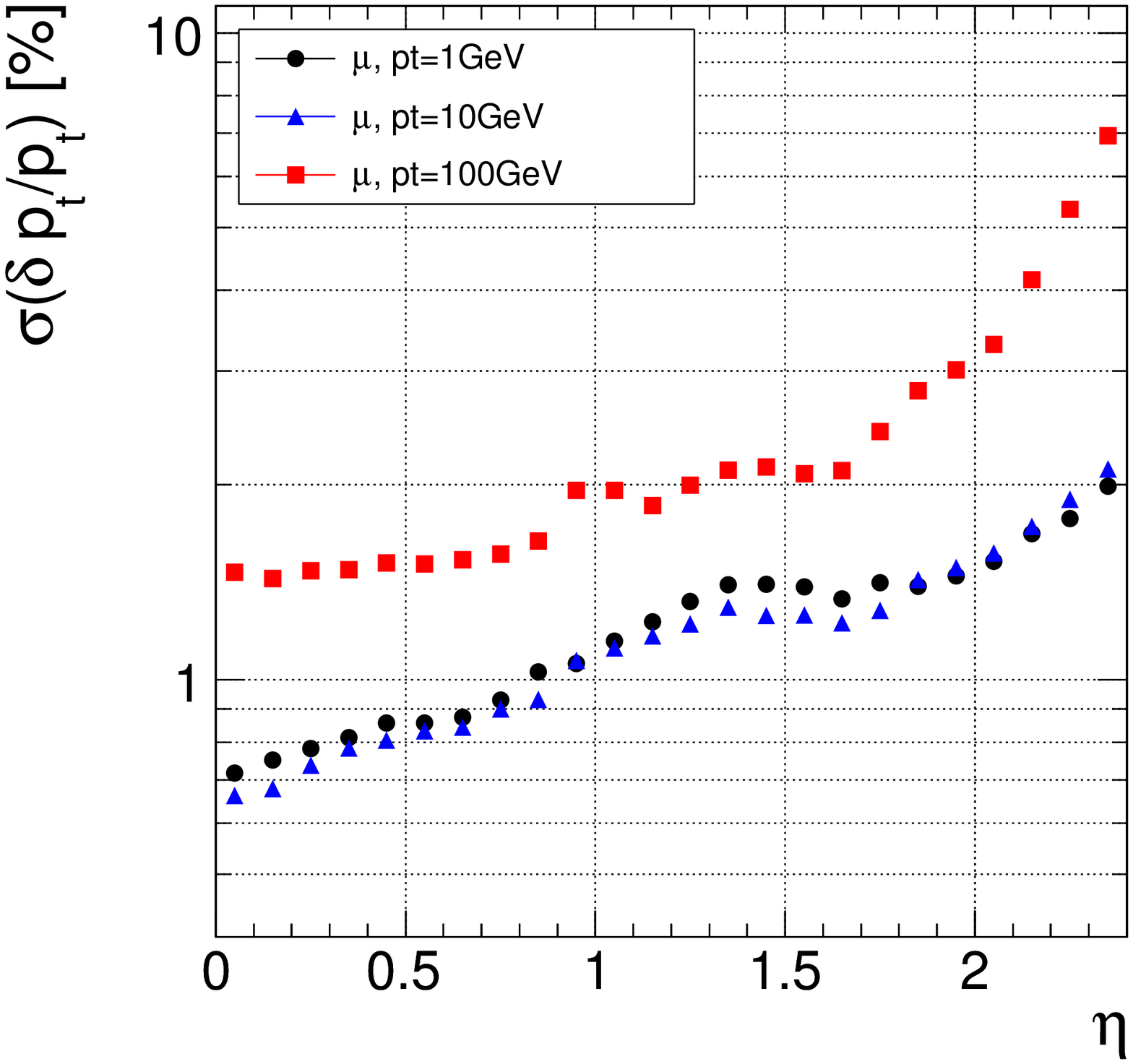}
\caption{
Resolution in transverse impact parameter $d_0$ (left) and in $p_T$ (right)
for muons with $p_T=1,10$ and $100 \rm\ GeV$.  }
\label{fig:resolution}
\end{figure*}

Five parameters are chosen to describe a track: The transverse and
longitudinal impact parameters $d_0$ and $z_0$, the angular parameters
$\phi$ and $\cot \theta$, and the transverse momentum $p_T$. The
resolutions in $d_0$ and in $p_T$ are shown in
Figure~\ref{fig:resolution}.

At high momentum, the impact parameter resolution is fairly constant
and is dominated by the hit resolution of the first hit in the pixel
detector. At lower momenta, the $d_0$ resolution is progressively
degraded by multiple scattering, until the latter becomes dominant.

The transverse momentum resolution is around $1\ldots2\%$ up to a
pseudorapidity of $|\eta|<1.6$ at high momentum.  For higher values of
$|\eta|$ the lever arm of the measurement is reduced. The degradation
around $|\eta|=1.0$ is due to the gap between the barrel and the
endcap disks.  At $p_T=100 \rm\ GeV$, the tracker material accounts
for $20\ldots30\%$ of the transverse momentum resolution. At lower
momenta, the resolution is dominated by multiple scattering and its
distribution reflects the amount of material traversed by the track.

\section{IMPACT OF MISALIGNMENT}
\label{sec:misali}

The large number of independent silicon sensors and their excellent
intrinsic resolution of $10\ldots50\rm\ \mu m$ make the alignment of
the CMS strip and pixel trackers a complex of challenging task. The
residual alignment uncertainties should not lead to a significant
degradation of the intrinsic tracker resolution.  For example, to
achieve a desired precision on the measurement of the $W$ boson mass
of $15\ldots20 \rm\ MeV$, the momentum scale has to be known to an
accuracy of $0.02$ to $0.025\%$, which implies the absolute detector
positions to be known with a precision of better than $10\rm\ \mu m$
in the $r\phi$ plane.
Misalignment will degrade the track parameter resolution and hence
affect the physics performance of the tracker, for instance the mass
resolution of resonances and b-tagging and vertexing performances.

In order to assess the impact of misalignment on the tracking and
vertexing performance in general, but also in specific physics
channels in particular, a realistic model of misalignment 
effects~\cite{misaliscen} has
been implemented in the standard CMS software, where the displacement
of detector modules is implemented at reconstruction level using a
dedicated software tool which is able to move and rotate all tracker
parts (individual sensors as well as composed structures such as whole
layers or disks). In addition, the position error assigned to a
reconstructed hit can be increased by adding an additional error that
reflects the size of the assumed misalignment (alignment position
error).

\begin{table*}
\vspace{0.3cm}
\centering
\begin{tabular}{l|cc|cccc}
\hline
 & \multicolumn{2}{c|}{Pixel} &  \multicolumn{4}{c}{Silicon Strip} \\
 &           &           &  Inner    &  Outer  &  Inner  &  \\
 &  Barrel   &   Endcap  &  Barrel   &  Barrel &  Disk   &  Endcap \\
\hline
 {First Data Taking Scenario} &     &       &       &       &       &       \\
 Modules                          & 13  &  2.5  &  200  &  100  &  100  &   50  \\
 Ladders/Rods/Rings/Petals        &  5  &  5    &  200  &  100  &  300  &  100  \\
\hline
 {Long Term Scenario}         &     &       &       &       &       &       \\
 Modules                          & 13  &  2.5  &  20   &  10   &  10   &   5   \\
 Ladders/Rods/Rings/Petals        &  5  &  5    &  20   &  10   &  30   &  10   \\
\hline
\end{tabular}
\vspace{0.3cm}

\caption
{Mounting precisions (in ${\mu m}$) used in the misalignment
simulation.}
\label{tab:misalisim}

\end{table*}

Two default {\em misalignment scenarios} have been implemented in the software:

\begin{itemize}

\item {\bf First Data Taking Scenario:}
This scenario is supposed to resemble the expected conditions during
the first data taking of CMS (few $100 \rm\ pb^{-1}$ of accumulated
luminosity). It assumes that the pixel detector has been aligned to a
reasonable level using tracks. For the strip detector it is assumed
that no track-based alignment is possible due to insufficient high
$p_T$ track statistics, so that only survey information is
available. In addition, the LAS would provide constraints on the
positions of the larger structures of the strip tracker.

\item {\bf Long Term Scenario: }
It is assumed that after the first few $\rm\ fb^{-1}$ of data have
been accumulated, a first complete track-based alignment down to the
sensor level has been carried out, resulting in an overall alignment
uncertainty of the strip tracker of $\sim 20\rm\ \mu m$.

\end{itemize}

\begin{figure*}
\centering
\includegraphics[width=0.49\linewidth]{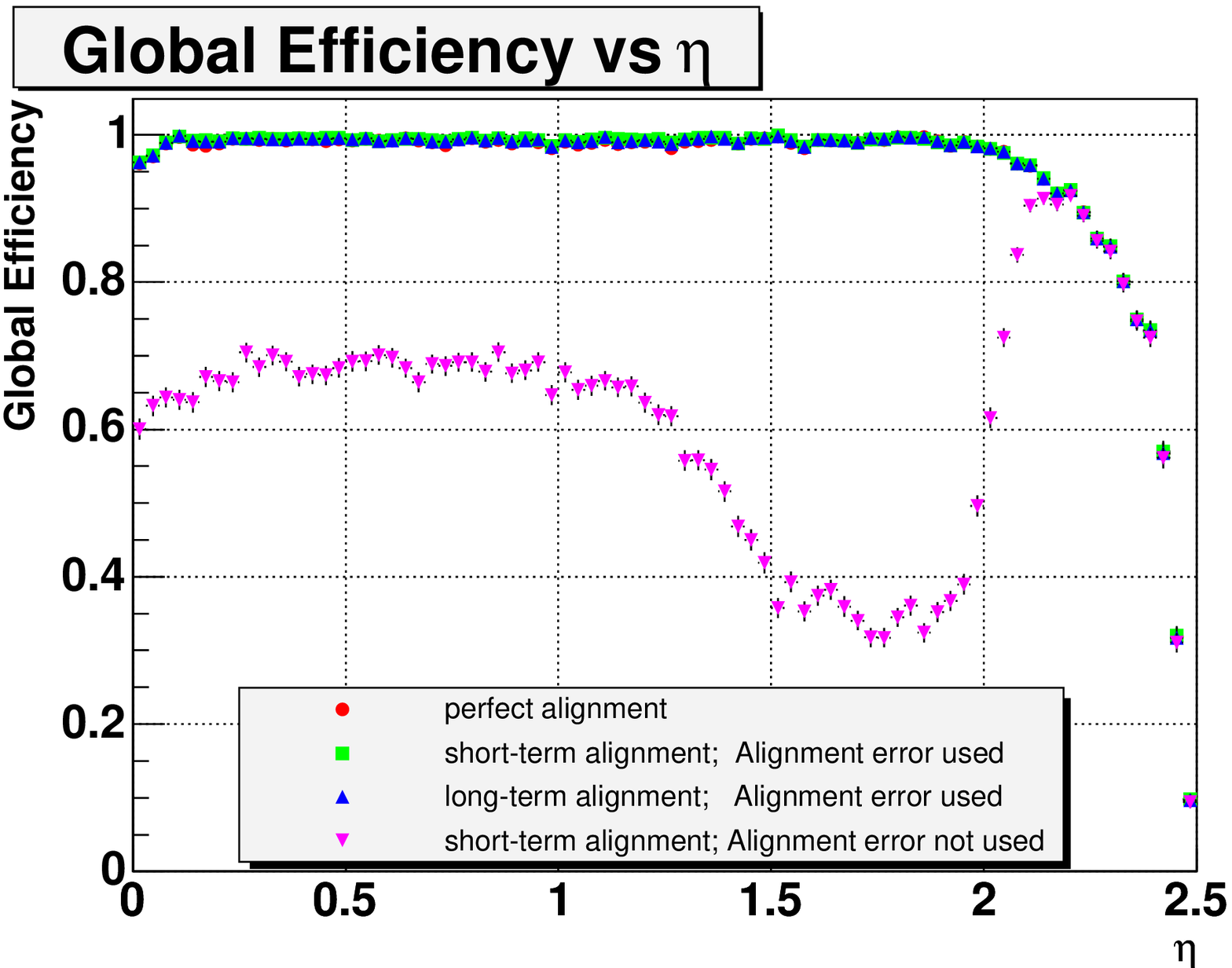}
\includegraphics[width=0.49\linewidth]{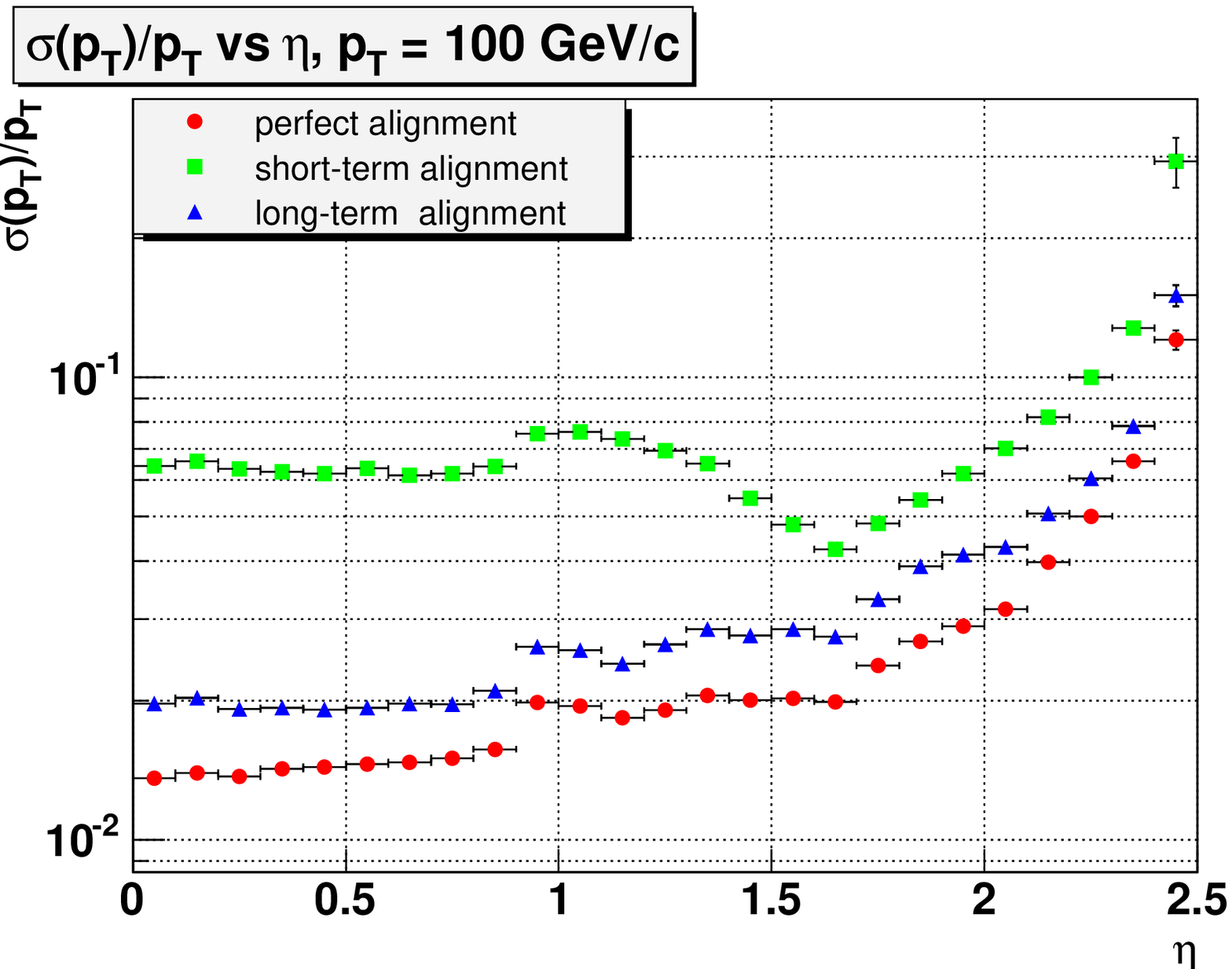}
\caption{
Track finding efficiency (left) and pt resolution vs $\eta$ (right)
for muons with $p_T=100 \rm\ GeV$. If the alignment uncertainty is not
accounted for, the efficiency is significantly degraded. The $p_T$
resolution deteriorates significantly with misalignment, in particular
for the short-term scenario.
}
\label{fig:misalignment}
\end{figure*}

The placement uncertainties used in the scenarios are listed in
Table~\ref{tab:misalisim}.  As an illustration of the implementation
and use of these misalignment scenarios, Figure~\ref{fig:misalignment}
shows the effects of misalignment on track-finding efficiency and
transverse momentum resolution for single muons \cite{misalinote}.
The track finding efficiency is close to unity for $|\eta|<2$ for all
misalignment scenarios, provided the alignment position error is taken
into account.  If not, the efficiency is significantly reduced, which
is illustrated in Figure~\ref{fig:misalignment} for the short term
scenario. The dip in the distribution in the range $1.2<|\eta|<2.0$ is
due to tracks passing through the TID, which has large alignment
uncertainties due to the missing laser alignment system. 
For $|\eta|>2.2$ the inclusion
of the alignment position error does not improve the efficiency due to
the large track extrapolation uncertainties involved in the very
forward direction.


\section{ALIGNMENT OF THE CMS TRACKER}

The alignment strategy for the CMS tracker forsees that in addition to
the knowledge of the positions of the modules from measurements at
construction time, the alignment will proceed by two means: A Laser
Alignment System (LAS) and track-based alignment.

\subsection{Laser Alignment System}

\begin{figure}
\centering
\includegraphics[width=0.99\linewidth]{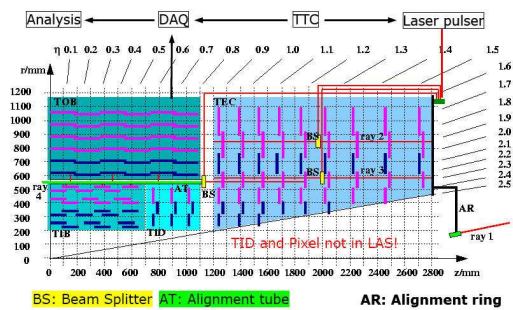}
\caption{Illustration of the CMS Laser Alignment System.
The laser beams are distributed by beam splitters (BS) and alignment
tubes (AT). The link to the muon system is implemented on the
alignment rings (AR) that are connected to the tracker back disks.  
}
\label{fig:las}
\end{figure}

The Laser Alignment System uses infrared laser beams to monitor the
positions of selected detector modules of the strip tracker and of
special alignment sensors in the muon system. Therefore it operates
globally on the larger tracker composite structures (TIB, TOB, TEC
disks) and cannot determine the position of individual modules.  The
goal of the LAS is to provide alignment information on a continuous
basis, providing position measurements of the tracker substructures at
the level of $100 \rm\ \mu m$, which is mandatory for pattern
recognition and for the High Level Trigger. In addition possible
structure movements can be monitored at the level of $10 \rm\ \mu m$.

The LAS design is illustrated in Figure~\ref{fig:las}.  Each tracker
endcap (TEC) uses in total 16 beams distributed in $\phi$ and crossing
all 9 TEC disks, which are used for the internal alignment of the TEC
disks.  The other 8 beams are foreseen to align TIB, TOB and TEC with
respect to each other. Finally, there is a link to the muon system.
As laser pulses are fired with a rate of around 100 Hz, a full snapshot
of the tracker structure can be taken in a few seconds.  The LAS is
foreseen to operate both in dedicated runs and during physics data
taking, so that the alignment can be monitored on a continuous basis.


\subsection{Track Based Alignment}

Track-based alignment was shown to be the optimal method for the
alignment of large tracking detectors in previous
experiments. However, it represents a major challenge at CMS because
the number of degrees of freedom involved is very large: Considering
3+3 translational and rotational degrees of freedom for each of the
$\sim 15000$ modules leads to $\mathcal{O}(100,000)$ alignment
parameters, which have to be determined with a precision of $\sim 10
\rm\ \mu m$. Moreover, the full covariance matrix is of size
$\mathcal{O}(10^{10})$.

In CMS, three different track-based alignment algorithms are
considered, some having been established at other experiments, others
newly developed. In the following, the main features and initial
results of using these algorithms in CMS are summarized.

\subsubsection{Kalman Filter}

\begin{figure*}
\centering
\includegraphics[width=0.45\linewidth]{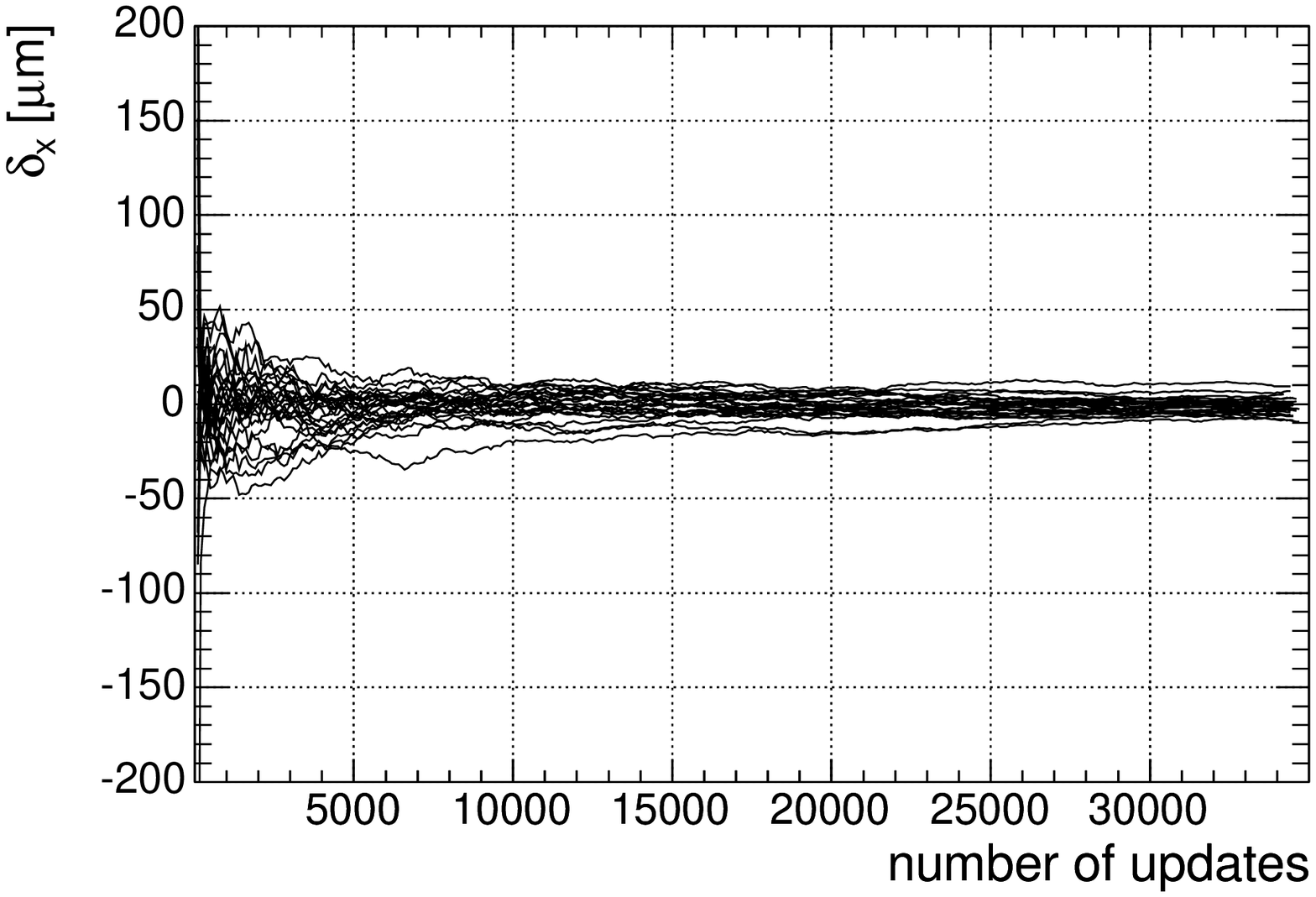}
\includegraphics[width=0.45\linewidth]{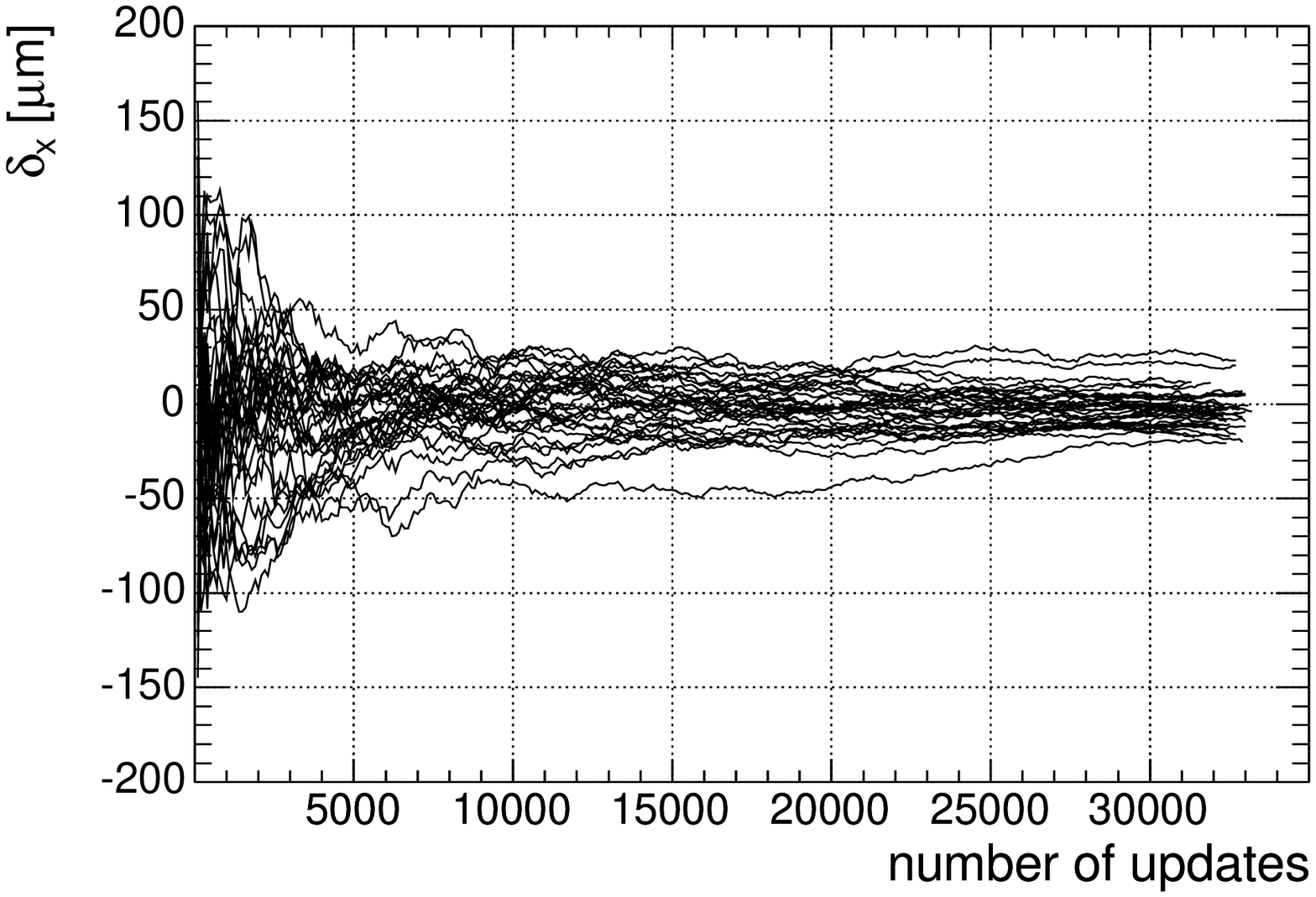}
\caption{
Kalman Filter alignment: Residuals in local x for TIB layers 1 (left)
and 2 (right) as a function of the number of processed tracks. }
\label{fig:kalman}
\end{figure*}

A method for global alignment using charged tracks can be derived from
the Kalman filter. The method is iterative, so that the alignment
parameters are updated after each track. It can be formulated in such
a way that no large matrices have to be inverted~\cite{kalmannote}. In
order to achieve a global alignment the update is not restricted to
the detector elements that are crossed by the track, but can be
extended to those elements that have significant correlations with the
ones in the current track. This requires some bookkeeping, but keeps
the computational load to an acceptable level.
It is possible to use prior information about the alignment obtained
from mechanical survey measurements as well as from laser
alignment. The algorithm can also be extended to deal with
kinematically constrained track pairs (originating from particle
decays).

The algorithm has been implemented in the CMS software and studied in
two small subsets of the silicon tracker: A telescope-like section of
the inner and outer barrel, and a wheel-like subset of the inner
barrel, consisting of 156 modules in 4 layers. The tracks used were
simulated single muons with $p_T=100 \rm\ GeV$. Random misalignment with a
standard deviation of $\sigma=100 \rm\ \mu m$ was applied to the local
$x$ and $y$ positions of the modules.  Results from the alignment of
the wheel-like setup are shown in Figure~\ref{fig:kalman}.  It shows
the evolution of the differences between true and estimated $x$-shifts
for layers 1 and 2. A total of 100 000 tracks were processed. As can
be seen, the speed of convergence depends on the layer. More results
can be found in~\cite{kalmannote}.

\subsubsection{Millepede-II}

Millepede~\cite{blobel} is a well established and robust program
package for alignment which has been used successfully at other
experiments, for example at H1, CDF, LHCb and others.  Being a
non-iterative method, it has been shown that it can improve the
alignment precision considerably with respect to other algorithms.

\begin{figure*}
\centering
\includegraphics[width=0.45\linewidth]{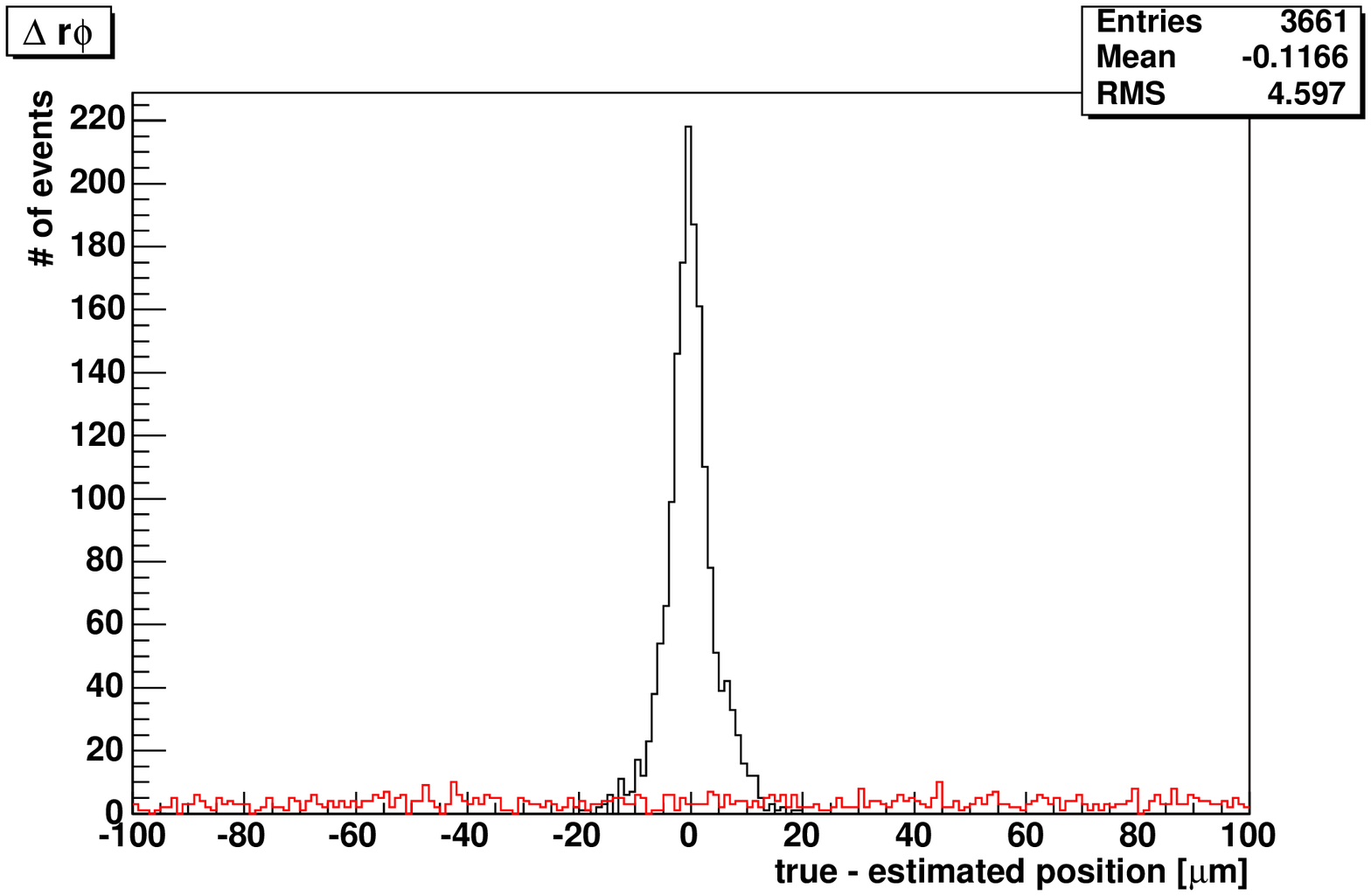}
\includegraphics[width=0.45\linewidth]{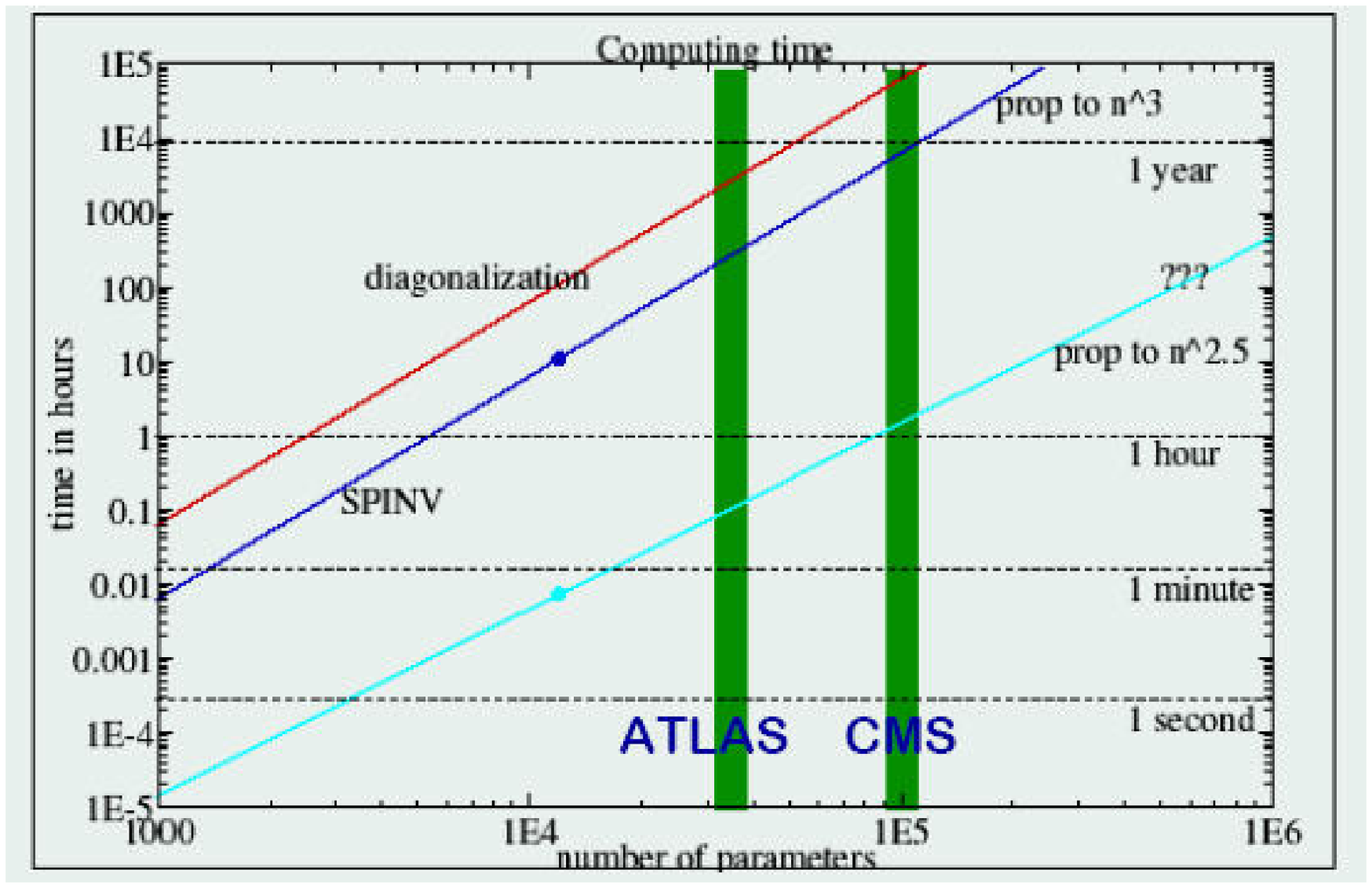}
\caption{
Millepede-II: Left: Residuals in $r\phi$ in the strip tracker barrel before (red) and
after (black) alignment using Millepede-II.
Right: CPU time as a function of alignment parameters for matrix
inversion (blue) and Millepede-II.
}
\label{fig:millepede}
\end{figure*}

Millepede is a linear least-squares algorithm which is fast, accurate
and can take into account correlations among parameters. In the
least-squares fit local track parameters and global alignment parameters
are fitted simultaneously.  The solution for the alignment parameters
is obtained from a matrix equation for the global parameters only.
For $N$ alignment parameters this requires the inversion of a $N {\rm
x} N$ matrix. However, this method can only be used up to $N\sim
10000$ due to CPU and memory constraints.  The alignment of the CMS tracker
exceeds this limit by one order of magnitude. Therefore, a new version
Millepede-II~\cite{millenote} was developed, which offers different
solution methods, and is applicable for $N$ much larger than $10000$.
In Millepede-II, in addition to the matrix inversion
and a diagonalization method, a new method for the solution
of very large matrix equations is implemented. This minimum
residual method applicable for sparse matrices determines
a good solution by iteration in acceptable time even for large $N$.

Millepede-II has been interfaced to the CMS software and the
alignment of parts of the CMS tracker has been carried out using
different scenarios~\cite{millenote}.  As an example,
Figure~\ref{fig:millepede} (left) shows hit residuals in $r\phi$ for
the new iterative method. Each individual sensor of the tracker was
misaligned.  The alignment procedure was carried out in the barrel
region ($|\eta|<0.9$) of the strip tracker using 1.8 million
$Z^0\rightarrow\mu^+\mu^-$ events.  The pixel layers and the outermost
barrel layer were kept fixed, resulting in $\sim 8400$ alignment
parameters. The convergence is very good, and the results obtained are
identical to those using the matrix inversion method, but the new
method being faster by about three orders of magnitude.

Figure~\ref{fig:millepede} (right) shows the needed CPU time as a
function of the number of alignment parameters for the diagonalization
and matrix inversion methods as well as for the new method used in
Millepede-II.  It can be seen that Millepede-II is expected to be
capable to solve the full CMS tracker alignment problem within
reasonable CPU time.

\subsubsection{HIP Algorithm}

\begin{figure*}
\centering
\includegraphics[angle=270,width=0.9\linewidth]{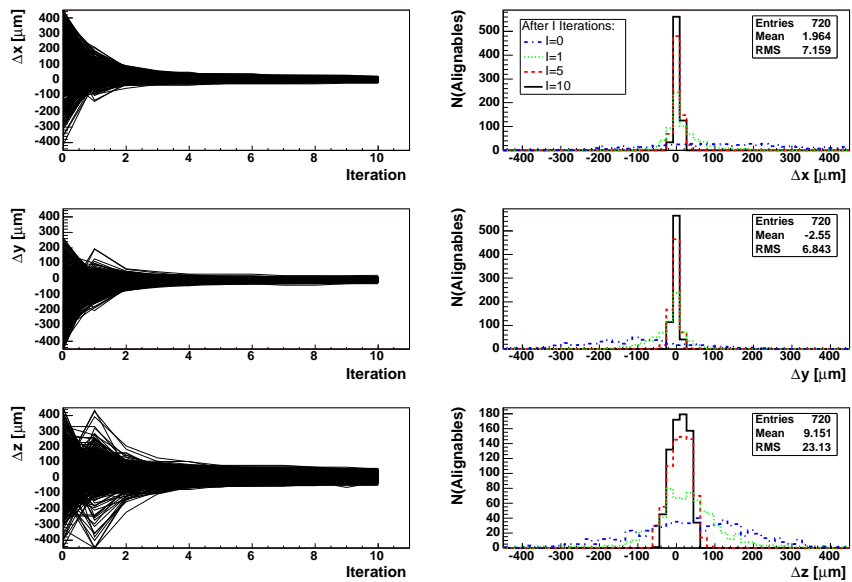}
\caption{
Alignment of the Pixel barrel modules with the HIP algorithm.
The residuals in global coordinates are shown as a function
of iteration (left) and projected for 0,1,5 and 10 iterations (right).
}
\label{fig:hip}
\end{figure*}

An iterative alignment algorithm using the Hits and Impact Points
(HIP) method was developed in~\cite{hiplajolla}. It is able to
determine the alignment of individual sensors by minimizing a local
$\chi^2$ function depending on the alignment parameters, constructed
from the track-hit residuals on the sensor.  Correlations between
different sensors are not explicitly included, but taken care of
implicitly by iterating the method, which involves consecutive cycles
of calculating the alignment parameters and refitting the tracks. The
algorithm is computationally light because no inversion of large
matrices is involved.
An alternative implementation of the algorithm is designed to align
composite detector structures for a common translation and
rotation~\cite{hipnote}, for example pixel ladders or layers. The
composite alignment involves only a small number of parameters, and
therefore a rather small number of tracks is sufficient to carry out
alignment already in the beginning of data taking.

The HIP algorithm has been used in~\cite{hipnote} for the alignment of
the pixel barrel modules using the First Data Taking misalignment
scenario (see section~\ref{sec:misali}).  The pixel endcaps and the
strip tracker are not misaligned.  The procedure has been iterated 10
times using 200 000 simulated $Z^0\rightarrow\mu^+\mu^-$
events. Figure~\ref{fig:hip} shows the differences between the true
and estimated alignment parameters. The convergence is good, with RMS
values of $7(23)\rm\ \mu m$ for the $x,y(z)$ coordinates,
respectively. The algorithm was also applied to a test beam
setup~\cite{hipcrack}.


\section{CONCLUSIONS}

The CMS silicon tracker is a complex device, consisting of more than
15000 individual silicon sensors. The track reconstruction performance
is very good, although the track reconstruction efficiency for low
momentum charged hadrons is affected by the significant amount of
tracker material.

Alignment of the tracker is a challenging task, and involves a laser
alignment system as well as track-based alignment with the goal to
determine the positions of all detector modules with a precision of
$10\rm\ \mu m$, so that the intrinsic resolution of the silicon
modules is not significantly degraded. To achieve this goal, CMS has
implemented three different track-based alignment algorithms.
Results from first alignment studies applying these algorithms to
parts of the CMS tracker in simulation are very encouraging.



\end{document}